\documentclass[aps,prb,twocolumn,showpacs,groupedaddress]{revtex4}
\usepackage{graphicx}

\begin{document}

\title{Large Magneto-Dielectric Effects in Orthorhombic HoMnO$_{3}$ and YMnO$%
_{3}$}
\author{B. Lorenz$^{1}$, Y. Q. Wang$^{1}$, Y. Y. Sun$^{1}$, and C. W. Chu$%
^{1,2,3}$} \affiliation{$^{1}$Department of Physics and TCSUH,
University of Houston, Houston, TX 77204-5002}
\affiliation{$^{2}$Lawrence Berkeley National Laboratory, 1
Cyclotron Road, Berkeley, CA 94720} \affiliation{$^{3}$Hong Kong
University of Science and Technology, Hong Kong, China}

\begin{abstract}
We have found a remarkable increase (up to 60 \%) of the dielectric
constant with the onset of magnetic order at 42 K in the metastable
orthorhombic structures of YMnO$_{3}$ and HoMnO$_{3}$ that proves
the existence of a strong magneto-dielectric coupling in the
compounds. Magnetic, dielectric, and thermodynamic properties show
distinct anomalies at the onset of the incommensurate magnetic order
and thermal hysteresis effects are observed around the lock-in
transition temperature at which the incommensurate magnetic order
locks into a temperature independent wave vector. The Mn$^{3+}$
spins and Ho$^{3+}$ moments both contribute to the
magneto-dielectric coupling. A large magneto-dielectric effect was
observed in HoMnO$_{3}$ at low temperature where the dielectric
constant can be tuned by an external magnetic field resulting in a
decrease of up to 8 \% at 7 Tesla. By comparing data for YMnO$_{3}$
and HoMnO$_{3}$ the contributions to the coupling between the
dielectric response and Mn and Ho magnetic moments are separated.
\end{abstract}

\maketitle











The coupling between dielectric and magnetic properties recently
observed in some manganites\cite{1,2,3,4,5,6} and in other
oxides\cite{7,8} is of fundamental interest and of eminent
significance for potential applications. The anomalies of dielectric
(magnetic) properties at magnetic (ferroelectric) phase transitions
and the possibility of tuning the dielectric constant
(magnetization) by external magnetic (electric) fields open new
perspectives in the basic understanding of the interesting materials
and for the design of new devices. The magneto-dielectric effect can
be explained by a spin-lattice coupling due to an increase of
magnetic exchange energy when the magnetic ions shift their
positions.\cite{9,10} This effect is particularly strong close to or
below a magnetic phase transition and may result in structural
anomalies and a change of the dielectric properties.

The rare earth manganites, $R$MnO$_{3}$ ($R$=rare earth metal), exhibit
strong magnetic exchange interactions between the magnetic moments of the Mn$%
^{3+}$ ions as well as some of the magnetic $R^{3+}$. Depending on the rare earth ionic size, $R$%
MnO$_{3}$ crystallizes in either hexagonal or orthorhombic
(distorted perovskite) structure with the structural phase boundary
between Ho and Dy. However, some of the hexagonal compounds can also
be synthesized as a metastable phase in the orthorhombic structure
by either special chemical procedures\cite{11} or high pressure
synthesis.\cite{12} The hexagonal phases of $R$MnO$_{3}$ show
ferroelectricity below Curie temperatures between 590 and 1000 K and
antiferromagnetic (AFM) transitions below 100 K with small but
distinct anomalies in the dielectric constant at or below the
magnetic transitions.\cite{1,2,3,4} Since symmetry arguments do not
allow a direct coupling between the Mn magnetic order and the
polarization in the hexagonal structure the observed
magneto-dielectric coupling has to be related to secondary
interactions but a microscopic explanation of these effects is not
yet available. Some work has been done to investigate possible
magneto-dielectric effects in the orthorhombic phases of $R$MnO$_{3}
$. Most notably is the recent discovery of a giant magnetoelectric
effect and the onset of ferroelectricity at the
incommensurate-commensurate AFM lock-in transition in
TbMnO$_{3}$.\cite{5} The report reveals a wealth of physical
phenomena in the spin-frustrated compound with different magnetic
ions and magnetoelastic as well as magnetoelectric coupling effects.

Spin frustration and incommensurate (IC) magnetic orders are typical
features of orthorhombic $R$MnO$_{3}$ with an Mn-O-Mn bond angle close to 145%
${{}^\circ}$%
. They are due to a competition of ferromagnetic and AFM
interactions between the Mn$^{3+}$ ions on nearest and next-nearest
neighbor positions mediated by the superexchange mechanism. The IC
AFM order is stable below about 50 K for rare earth ions from Eu to
Ho followed at lower temperatures by a transition into A-type AFM
(Eu, Gd), E-type AFM (Ho), or an IC magnetic structure with a fixed
modulation wave vector (Tb, Dy).\cite{13} Among the possible
candidates for further investigations YMnO$_{3}$ and HoMnO$_{3}$ are
of preferred interest. Both compounds exist in hexagonal and
orthorhombic structures. This allows a direct comparison of their
properties in different lattice symmetries. Furthermore, the ionic
radii of Y$^{3+}$ and Ho$^{3+}$ are very close so that the
structural parameters are almost identical. The Y$^{3+}$ ion is
nonmagnetic but the Ho$^{3+}$ carries a magnetic moment. The effect
of the additional magnetic species in HoMnO$_{3}$ can, therefore, be
resolved by comparing its properties with those of YMnO$_{3}$. Both
compounds exhibit an IC AFM transition at about 42 K (order of the
Mn$^{3+}$ spins) and a lock-in transition into a temperature
independent wave vector at lower T.\cite{14,15} HoMnO$_{3}$ shows
another transition below 9 K that was attributed to the AFM order of
the Ho$^{3+}$ moments.\cite{14} It is of primary interest to
investigate the dielectric constant and the magneto-dielectric
couplings at these transitions and to compare the results with
similar observations in the hexagonal $R$MnO$_{3}$.\cite{1,2}

We have therefore focused our attention onto the dielectric
properties of both compounds in the orthorhombic structure close to
the magnetic phase transitions and their dependence on external
magnetic fields. We have found a large magneto-dielectric effect
resulting in an enhancement of the dielectric constant, $\varepsilon
$, at zero magnetic
field below the IC AFM transition by 60 \% in YMnO$_{3}$ and 42 \% in HoMnO$%
_{3}$, respectively. In HoMnO$_{3}$ we also show a strong dependence of $%
\varepsilon $ on external magnetic fields below the lock-in transition from
the IC to the commensurate magnetic order.

Single-phase hexagonal samples with nominal composition YMnO$_{3}$ (HoMnO$%
_{3}$) were prepared by a solid-state reaction technique. Prescribed amounts
of Y$_{2}$O$_{3}$ (Ho$_{2}$O$_{3}$) and Mn$_{2}$O$_{3}$ were mixed,
preheated at 900$^{o}$ C (1000$^{o}$ C) in O$_{2}$ for 16 hours, and
sintered at 1150$^{o}$ C (1100$^{o}$ C) for 24 hours under an oxygen
atmosphere. The hexagonal compounds were transformed into the orthorhombic
structure by high-pressure sintering for 5 hours (1020$^{o}$ C, 3.5 GPa).
The phase pure Pbnm orthorhombic structure was obtained and no impurity
phases could be detected in the x-ray spectra for both compounds. The
samples were shaped for dielectric measurements into pellets about 0.5 mm
thick with a contact area of 10 mm$^{2}$. The capacitance was measured
between 100 kHz and 1 MHz using the HP 4285A meter and the samples were
exposed to magnetic fields up to 7 Tesla in the Physical Property
Measurement System. Magnetization measurements were conducted in fields up
to 5 Tesla employing the Magnetic Property Measurement System.

The temperature dependence of the dielectric constant at 100 kHz is
shown below 60 K for YMnO$_3$ and HoMnO$_3$ in Fig. 1 and Fig. 2,
respectively. Below the IC Neel temperature, T$ _{N}$, $\varepsilon
$(T) increases rapidly and passes through a maximum at lower T. The
enhancement of $\varepsilon $ is more than 60 \% in YMnO$_{3}$ and
42 \% in HoMnO$_{3}$. The Neel temperatures for both compounds are
nearly identical, T$_{N}$=42.2 K, due to the structural similarity
of both compounds. Besides the large increase of $ \varepsilon $(T)
there is also a pronounced thermal hysteresis with decreasing and
increasing temperatures well below T$_{N}$. Whereas T$_{N}$
is exactly the same upon cooling and heating the hysteresis of $\varepsilon $%
(T) is essential at temperatures below about 30 K. Therefore, it
cannot be attributed to the IC Neel transition but it is rather
related to the lock-in transition at which the modulation vector of
the AFM order of the Mn$^{3+}$ spins locks into a T-independent
value.\cite{14,15} The lock-in transition temperature, T$_{L}$, was
estimated from neutron scattering data as 28 K
and 26 K for YMnO$_{3}$ and HoMnO$_{3}$, respectively. While the maxima of $%
\varepsilon $(T) appear at lower T the largest slope of the increasing $%
\varepsilon $(T) below T$_{N}$ is close to the T$_{L}$ values given
above. For YMnO$_{3}$ the steepest increase of $\varepsilon $(T) is
at 27.5 and 29.6 K with decreasing and increasing T, respectively.
The corresponding values for HoMnO$_{3}$ are 23 and 26 K. Both sets
of critical temperatures are very close to the T$_{L}$'s from
neutron scattering. Therefore, we associate the temperatures of the
steepest increase of $\varepsilon $(T) with T$_{L}$ and we conclude
that the magnetic lock-in transitions show a thermal hysteresis of
about 2 to 3 K that is typical for first order phase transitions.
Similar hysteresis effects have also been reported in other rare
earth manganites very recently.\cite{16}

The distinct anomalies in the dielectric constant and its closeness
to the magnetic transitions suggest a very strong magneto-dielectric
coupling. In order to establish the correlation of the magnetic
order and the dielectric anomalies and, in particular, the thermal
hysteresis observed at T$_{L}$, we have measured the dc
magnetization of YMnO$_{3}$ and HoMnO$_{3}$ between 2 and 400 K. The
high temperature data show the characteristic Curie-Weiss behavior
with an extrapolated paramagnetic temperature of -54 K and -19 K as
well as an effective magnetic moment of 5.0 $\mu _{B}$ and 10.8 $\mu
_{B}$ for YMnO$_{3}$ and HoMnO$_{3}$, respectively. The sets of
Curie-Weiss parameters are in good agreement with previous
reports.\cite{14,15,17} The effective moments are close to the
theoretical values of 4.9 $\mu _{B}$ for YMnO$_{3}$ and 11.5 $\mu
_{B}$ for HoMnO$_{3}$. In the low temperature range the IC magnetic transition of YMnO$%
_{3}$ is clearly indicated by the maximum of the dc susceptibility
(Fig. 1) at T$_{N}$=42.2 K which coincides with the critical
temperature deduced from the upturn of $\varepsilon $(T) discussed
above. The pronounced shoulder of the susceptibility at lower T
followed by a fast decrease defines the lock-in transition. The
distinct hysteresis of the susceptibility close to the lock-in
transition with the critical temperatures of 27.4 K upon cooling and
28.7 K upon heating has not been observed before but it is in
perfect agreement with the dielectric data. The observation of the
magnetic hysteresis (Fig. 1, upper curve) provides further evidence
for the first order nature of this transition and unambiguously
proves the coupling between magnetic order and dielectric
properties. The dc magnetization of HoMnO$_{3}$ is dominated by the
large paramagnetic moment of the Ho$^{3+}$. This contribution
increases the magnetization of HoMnO$_{3}$ in the low-T range by a
factor of up to 20 as compared to YMnO$_{3}$. Therefore, anomalies
at the IC spin order transition of the Mn$^{3+}$ are barely detected
in the data of Fig. 2 (upper curve). However, when the inverse
susceptibility is differentiated with respect to T a small but
distinct peak appears at $T_{N}$=42.2 K. The major anomaly in the
HoMnO$_{3}$ magnetization is the peak below 10 K that is due to the
AFM order of the Ho$^{3+}$ moments. Magnetic hysteresis effects as
revealed in the $\varepsilon $(T) data are barely seen in Fig. 2
since the subtle changes in the Mn$^{3+}$ spin order at T$_{L}$ are
concealed by the huge contribution of the paramagnetic Ho$^{3+}$.
Neutron scattering experiments have shown a similar hysteresis in
the magnetic reflections between 10 K and 35 K in
HoMnO$_3$.\cite{17} We have measured the heat capacity, C$_{p}$(T),
of both compounds and observed small hysteresis effects proving the
thermodynamic origin of the thermal hysteresis as seen in the
dielectric and the magnetic (YMnO$_{3}$) data. In HoMnO$_{3}$ the
C$_{p}$(T) for cooling and heating cycles differs by up to 200
mJ/(mol K) in the temperature range between 10 K and 35 K in which
also the hysteresis of $\varepsilon $(T) was observed (Fig. 2). In
YMnO$_{3}$ the C$_{p}$(T) of the heating cycle is up to 300 mJ/mol K
enhanced with respect to the cooling data between 23 and 35 K, close
to the interval of magnetic hysteresis (Fig. 1). Therefore, we
conclude that thermal hysteresis below T$_{N}$ is an intrinsic
property for both compounds and that it is related to the
development of the IC magnetic order and the lock-in transition to a
T-independent wavelength. The hysteresis effects on the
magnetization and specific heat are very small but they appear far
more pronounced in $\varepsilon $(T). The dielectric constant is
extremely sensitive to subtle changes of the magnetic order and
serves as a perfect probe of the magnetic state.

The IC AFM order of the Mn$^{3+}$ is very rigid with respect to
external magnetic fields. The dc susceptibilities of YMnO$_{3}$
measured at 1 T and at 5 T, for example, are basically identical to
the low-field data shown in Fig. 1 over the whole temperature range.
No shift of T$_{N}$ was observed
with increasing H. In HoMnO$_{3}$, however, the magnetic order of the Ho$%
^{3+}$ moments at low T is rapidly suppressed by the magnetic field.\cite%
{14} The field dependence of the magnetization (Fig. 2, inset)
reveals a metamagnetic transition below the Ho-ordering temperature
as indicated by the maximum of M/H at $H_m$=1.5 Tesla (6 K) and the
hysteresis of M/H at low fields. The metamagnetic transition,
observed also in recent measurements at low T,\cite{12,14}
disappears for $T\approx$9 K, the transition temperature of the AFM
Ho ordering.

The magnetic tunability of dielectric properties is of
particular interest. The magnetic field dependence of the dielectric constant of YMnO$%
_{3}$ is small as shown in Fig. 3 (only cooling data are included in
Fig. 3). In HoMnO$_{3}$, however, the external field causes a
sizable decrease of $\varepsilon $(T) in the low-T range, the
magneto-dielectric effect at 4.5 K is shown in the inset of Fig. 3.
The dielectric constant decreases by almost 8 \% at 7 Tesla. This
large magneto-dielectric effect provides evidence for a strong
coupling of the dielectric response to the Ho$^{3+}$ magnetic
moments via magnetoelastic effects. The decrease of $\varepsilon
$(T) with H sets in below 22 K, the temperature of the lock-in
transition into the E-type commensurate magnetic order. The AFM
order of the Ho spins below 9 K reduces the magnetic field effect on
$\varepsilon $ at low fields and $\varepsilon $(H) is almost
constant for $H<H_m$. However, $\varepsilon $(H) decreases rapidly
at fields above the metamagnetic transition as shown in the inset of
Fig. 3. Neutron scattering experiments\cite{14} have reported a
small AFM magnetic moment of the Ho below 22 K that increases
sharply at the main transition close to 7 K. The Ho moments lie in
the (1,0,1) planes and their angle with the orthorhombic c-axis
increases suddenly at about 15 K
and moves continuously up to 60$%
{{}^\circ}%
$ at T=0.\cite{14} The sharp change of spin direction at 15 K is
reflected in the small but distinct drop of $\varepsilon $(T) right
below its maximum temperature (Fig. 2). The magnetic field component
in the (1,0,1) plane can rotate the Ho moments and, via
magnetoelastic effects, change the dielectric constant. Below the
AFM transition temperature (about 9 K) the Ho-spin system becomes
less susceptible to the external field and the magneto-dielectric
effect is small. Only when H increases above $H_m$ the Ho moments
respond to H by rotating in the a-c plane resulting in the observed
decrease of $\varepsilon $(H). We therefore conclude that the
observed magneto-dielectric effect in HoMnO$_3$ is a consequence of
the field-induced rotation of the Ho spins in the (1,0,1) plane. It
should be noted that all observations are made using polycrystalline
samples and all measured quantities are averaged over the random
grain orientation. It is expected that the magneto-dielectric effect
is even larger when well oriented single crystals could be
investigated. Unfortunately, single crystals of orthorhombic
HoMnO$_3$ grown under high-pressure conditions are not yet
available.

It is interesting to compare the large magneto-dielectric effect
observed in orthorhombic YMnO$_{3}$ and HoMnO$_{3}$ with similar
dielectric anomalies in the hexagonal structures of the same
compounds. The hexagonal manganites order in a frustrated AFM spin
arrangement at 71 K (YMnO$_{3}$) and at 76 K (HoMnO$_{3}$). Here
spin frustration is due to the geometric constraint in the
triangular lattice formed by the Mn-ions in the a-b plane. In
HoMnO$_{3}$ two additional magnetic transitions have been observed
at 33 K and 5 K related to Mn$^{3+}$-spin rotation and magnetic
ordering of Ho$^{3+}$ moments. Anomalies of the dielectric constant
are observed at all magnetic transitions in the hexagonal
(Y/Ho)MnO$_{3}$, however, the magnitude of these anomalies is
small.\cite{1,2,3} Even the recently reported sharp peak of
$\varepsilon $(T) at the spin-rotation transition of hexagonal
HoMnO$_{3}$, the strongest dielectric anomaly in hexagonal
RMnO$_{3}$, does not exceed in magnitude about 4 to 5 \% of the base
$\varepsilon $.\cite{2} The small response in the hexagonal
structures can be explained by the existing ferroelectric order that
forms well above room temperature in hexagonal YMnO$_3$ and
HoMnO$_3$. At the temperature of the magnetic transitions ($<$ 100
K) the electric polarization is rigid and any effects of magnetic
order or magnetic fields on the dielectric constant are therefore
small. The 60 \% increase of $\varepsilon $ below the AFM phase of
the orthorhombic structure is large and it indicates that huge
magneto-dielectric effects are expected in the orthorhombic
rare-earth manganites, as recently reported by Goto et al.\cite{16}

In summary, we have demonstrated the existence of a strong coupling between
dielectric properties and magnetic orders in the metastable orthorhombic
structures of YMnO$_{3}$ and HoMnO$_{3}$. The dielectric constant increases
by up to 60 \% in the IC magnetic phase below T$_{N}$=42.2 K. We show that
thermal hysteresis exists near the lock-in transition temperature, T$_{L}$,
in both compounds, typical for first-order phase transitions. Both, the Mn$%
^{3+}$ spins and the Ho$^{3+}$ moments, contribute to the increase
of $\varepsilon $(T). A large magneto-dielectric effect at low T in HoMnO$%
_{3}$ is explained by the magnetic field induced re-orientation of
the Ho moments in the a-c plane, resulting in a metamagnetic
transition below 9 K. By comparing data for YMnO$_{3}$ and HoMnO$%
_{3}$ the contributions to the coupling between the dielectric
response and Mn and Ho magnetic moments have been separated.


\begin{acknowledgments}
This work is supported in part by NSF Grant No. DMR-9804325, the
T.L.L. Temple Foundation, the John J. and Rebecca Moores Endowment,
and the State of Texas through the TCSUH at the University of
Houston and at Lawrence Berkeley Laboratory by the Director, Office
of Energy Research, Office of Basic Energy Sciences, Division of
Materials Sciences of the U.S. Department of Energy under Contract
No. DE-AC03-76SF00098.
\end{acknowledgments}


\begin{figure}[tbp]
\caption{Magnetization (circles, left scale) and dielectric constant
(squares, right scale) as function of temperature of orthorhombic
YMnO$_{3}$. Closed symbols: decreasing T; open symbols: increasing
T.}
\caption{Magnetization (circles, left scale) and dielectric constant
(squares, right scale) as function of temperature of orthorhombic
HoMnO$_{3}$. Closed symbols: decreasing T; open symbols: increasing
T. The inset shows the susceptibility vs. field at different
temperatures.}
\caption{Magneto-dielectric effect in YMnO$_{3}$ (left scale, full
line: H=0, dotted line: H=7 Tesla) and in HoMnO$_{3}$ (right scale,
full line: H=0, dotted lines: H=3,5,7 Tesla - from top to bottom).
Inset: Relative change of the dielectric constant as function of
magnetic field for HoMnO$_{3}$ at 4.5 K.}
\end{figure}

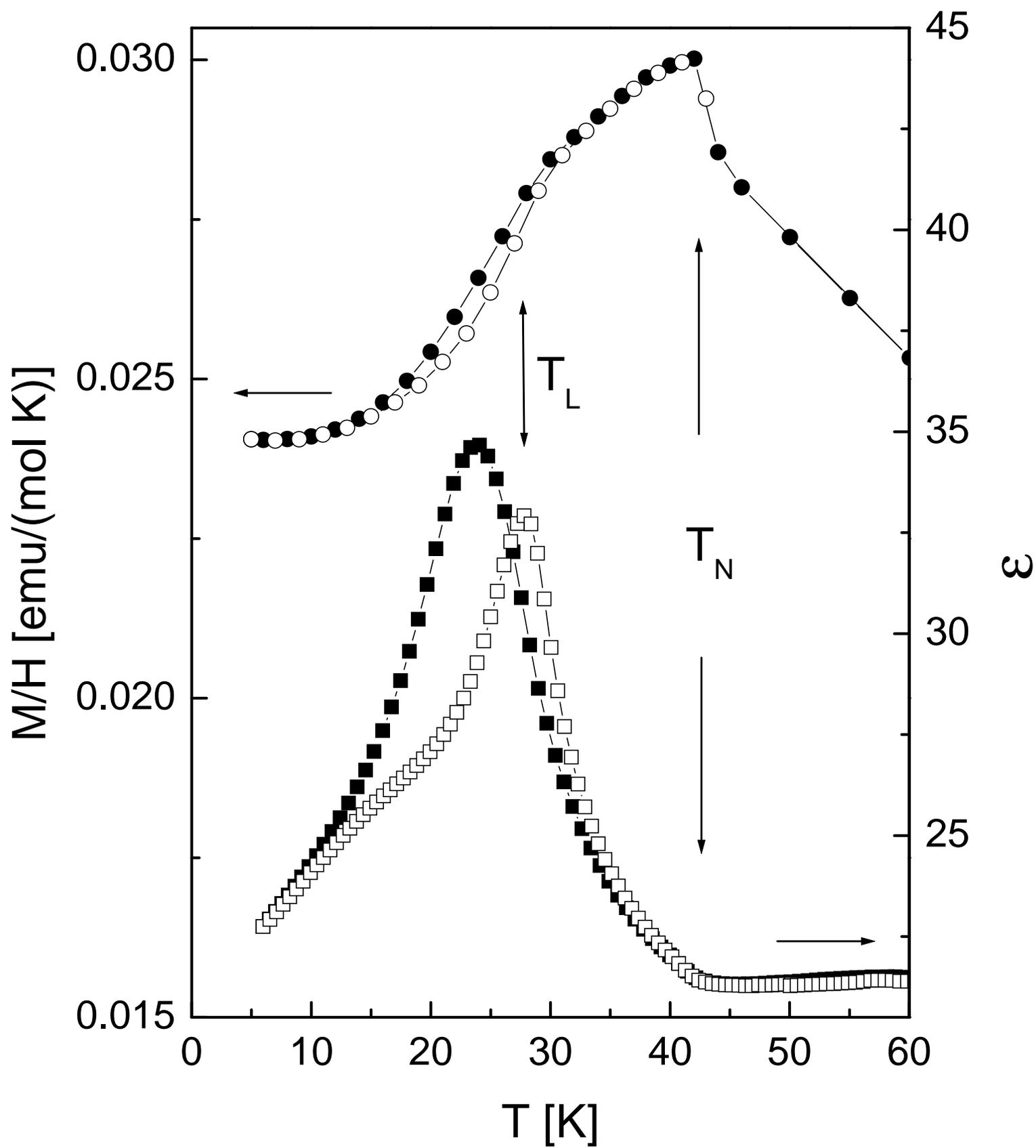
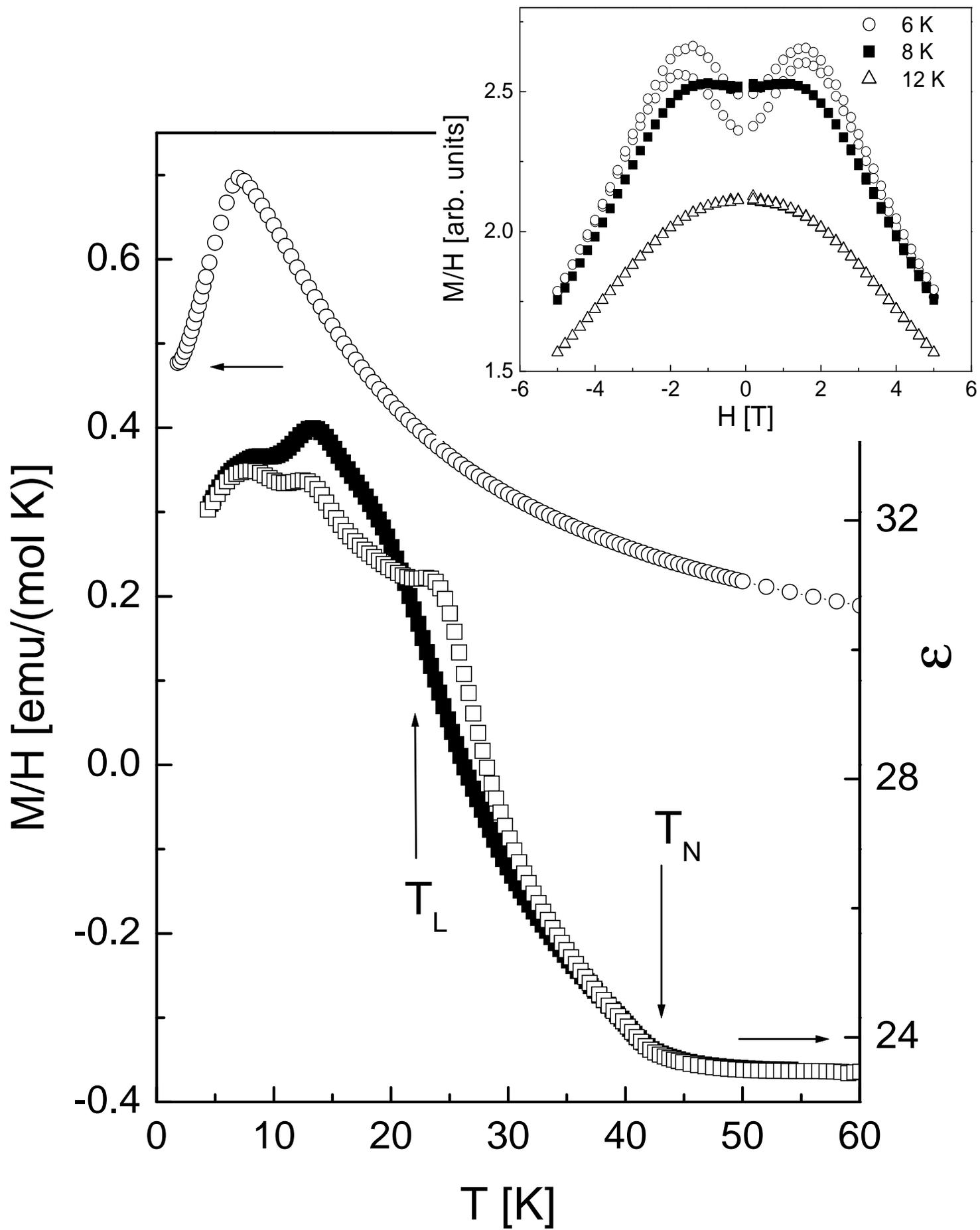
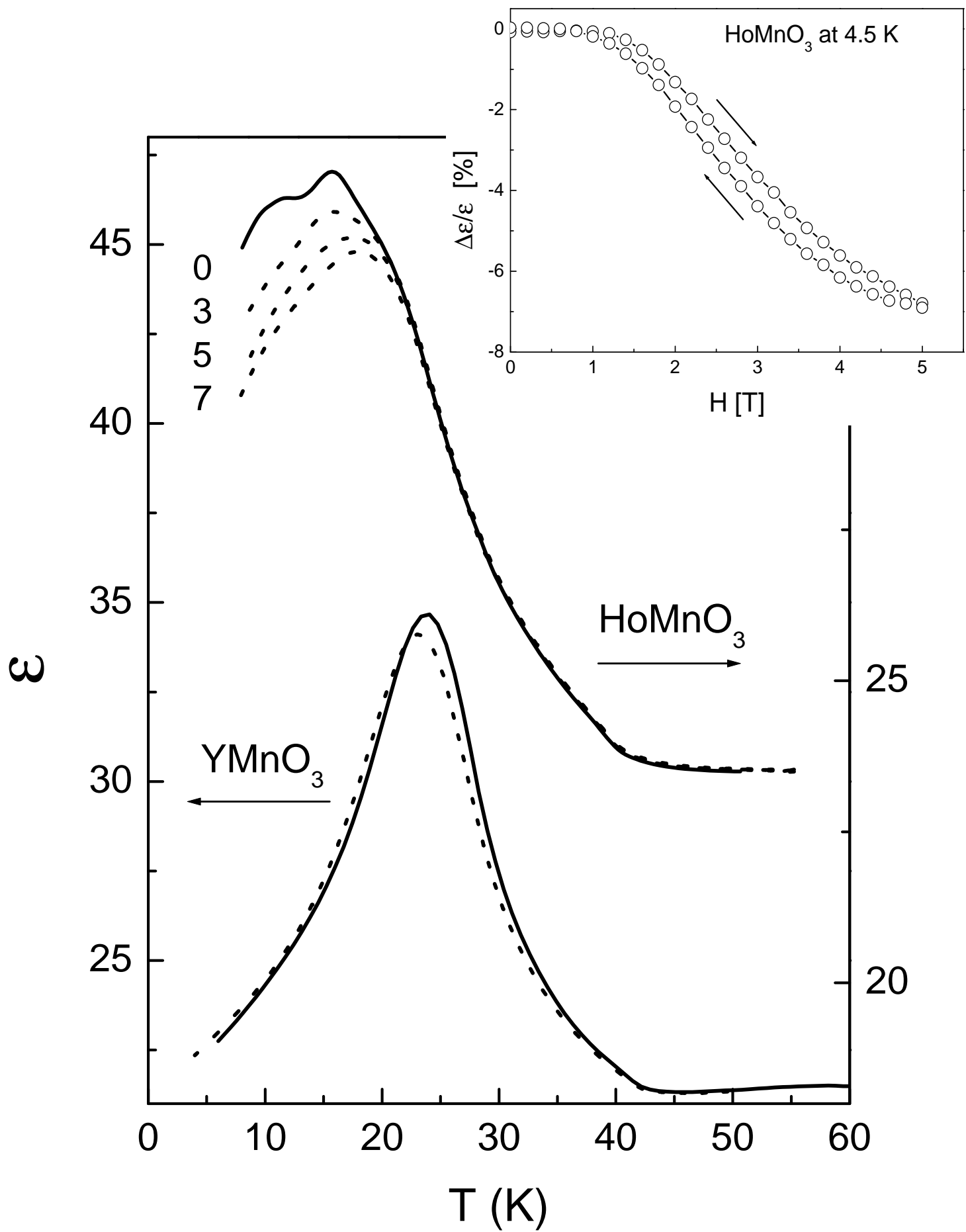
\end{document}